\documentclass{iopjournal}
\usepackage{amssymb}
\usepackage{amsmath}
\usepackage{lmodern}
\begin{document}

\articletype{Paper} %	 e.g. Paper, Letter, Topical Review...

\title{Non-Hermitian Exceptional Dynamics in First-Order Heat Transport}

\author{Pengfei Zhu$^{1,*}$\orcid{0000-0003-0431-1538}}

\affil{$^1$Bundesanstalt für Materialforschung and -prüfung (BAM), 12205 Berlin, Germany}

\affil{$^*$Author to whom any correspondence should be addressed.}

\email{pengfei.zhu@bam.de}


\begin{abstract}
Heat transport exhibits distinct regimes ranging from ballistic propagation to diffusive relaxation, traditionally described by disparate theoretical frameworks. Here, we introduce a unified first-order operator formulation in which temperature and heat flux are treated as a coupled state vector, yielding a minimal representative dynamical closure of heat transport. The resulting generator is intrinsically non-Hermitian and gives rise to a spectral structure governed by an exceptional point that separates overdamped diffusion from underdamped wave-like propagation. In this framework, Fourier’s law emerges as a singular limit of a hyperbolic dissipative system, while the Cattaneo equation arises naturally as the minimal representative hydrodynamic closure of kinetic theory. We show that the exceptional point induces nonanalytic spectral transitions, nonmodal transient dynamics, and a breakdown of conventional modal decomposition. The theory further generalizes to anisotropic media, where direction-dependent exceptional surfaces enable intrinsic steering of heat flow. Our results establish a unified non-Hermitian dynamical framework for heat transport and reveal exceptional-point physics as a fundamental organizing principle underlying thermal dynamics across scales.
\end{abstract}

\section{Introduction}
Heat transport spans multiple scales~\cite{ref1}, from the microscopic kinetic dynamics~\cite{ref2} to macroscopic diffusion~\cite{ref3}. At the kinetic level, transport is governed by the Boltzmann equation~\cite{ref4}; at the macroscopic level, it reduces to Fourier’s law~\cite{ref5}; and at intermediate scales, finite-speed propagation emerges through the Cattaneo equation~\cite{ref6}. Despite their apparent differences, these descriptions are not fundamentally distinct but reflect different limits of a unified dynamical structure. Guyer and Krumhansl developed a macroscopic phonon-transport theory based on eigenmode expansion of linearized Boltzmann equation~\cite{ref7}. Nevertheless, its predictive power is limited by parameter non-uniqueness and potential overfitting. Jou and Lebon~\cite{ref8} developed the extended irreversible thermodynamics (EIT), which promotes heat flux and other dissipative quantities to independent state variables, leading to first-order dynamical formulation. However, the choice of extended variables is not unique, and the resulting closure procedures rely on truncations of an underlying infinite hierarchy of moments. The Boltzmann kinetic theory~\cite{ref9} provides the most fundamental statistical description of heat transport. Its main difficulty lies in its high-dimensional phase-space structure and nonlinear collision integral. Fractional and memory-dependent heat transport models~\cite{ref10} generalize classical diffusion by incorporating long-time correlations and nonlocal temporal dynamics. However, the physical interpretation of fractional derivatives is not unique, and these models frequently function as effective fitting frameworks rather than predictive theories. Despite significant progress across these approaches, a fully unified dynamical description that simultaneously resolves the closure ambiguity~\cite{ref11}, preserves a clear dynamical structure~\cite{ref12}, and exposes the spectral nature of transport regimes remains elusive~\cite{ref13}. 

Here, we introduce a first-order operator formulation~\cite{ref14} of heat transport, in which temperature and heat flux are unified into an extended state vector. This formulation provides a physically constrained minimal representative of first-order heat transport dynamics~\cite{ref15} that naturally captures both diffusive and wave-like regimes, and reveals a non-Hermitian spectral structure~\cite{ref16} with an exceptional point (EP)~\cite{ref17} governing the transition between them. Here, minimality refers to the lowest-order extension consistent with conservation, finite propagation speed, isotropy, and dissipative stability, rather than strict uniqueness in an unconstrained operator space. The overdamped–underdamped transition in the Cattaneo equation is well known. The novelty here is not the transition itself, but its reinterpretation as a non‑Hermitian exceptional point of the extended transport generator. While the eigenvalue degeneracy coincides with the critical damping condition, the exceptional point is distinguished by the coalescence of eigenvectors and the resulting non‑diagonalizable generator, which has direct dynamical consequences in the time domain.

\section{Model}
We begin from the observation that classical diffusion arises as singular limit of a damped wave system~\cite{ref18} with dispersion relation:
\begin{equation}
\lambda^2 + \frac{1}{\tau}\lambda + c^2 k^2 = 0
\end{equation}
where $k$ is the wavevector characterizing spatial variation, and $\lambda$ is the (generally complex) temporal growth rate. The parameter $\tau$ denotes the relaxation time, and $c$ is the characteristic thermal propagation velocity defined by $c^2 = \frac{\alpha}{\tau}$  with $\alpha$ the thermal diffusivity. The real part of $\lambda$ determines decay, while its imaginary part corresponds to oscillatory behavior. In the singular limit $\tau \to 0$, the fast mode collapses and purely diffusive dynamics is recovered. This observation suggests that diffusion is not fundamental but rather emerges as a singular limit of a hyperbolic dissipative system~\cite{ref19}. Therefore, we introduce the extended state vector $\psi = (T, q)$, where $T(x,t)$ is the temperature field and $q(x,t) \in \mathbb{R}^d$ is the heat flux, and postulate a first-order evolution equation:
\begin{equation}
\partial_t \psi = -\mathcal{L}\psi
\end{equation}
with operator decomposition $\mathcal{L} = \mathcal{V} + \Gamma$, where
\[
\mathcal{V} =
\begin{pmatrix}
0 & \nabla \cdot \\
c^2 \nabla & 0
\end{pmatrix}, 
\qquad
\Gamma =
\begin{pmatrix}
0 & 0 \\
0 & \tau^{-1} I
\end{pmatrix}.
\]
Here, $\mathcal{V}$ is defined on the domain $H^1 \oplus H(\mathrm{div})$, ensuring that $\nabla T$ and $\nabla \cdot q$ are well-defined. This yields the explicit system:
\begin{equation}
\begin{cases}
\partial_t T + \nabla \cdot q = 0, \\
\tau \partial_t q + q = -c^2 \nabla T.
\end{cases}
\end{equation}
which recovers the Cattaneo-type constitutive relation in a dynamical form. The operator $\mathcal{V}$ exhibits an off-diagonal structure~\cite{ref20} that couples the conserved field $T$ to its conjugate flux $q$. In contrast to Hermitian wave operators~\cite{ref21}, the full generator $\mathcal{L}$ is intrinsically non-Hermitian due to the presence of the dissipative relaxation term $\Gamma$. Eliminating the flux variable yields a closed equation for the temperature field. Taking the time derivative of the conservation law gives
\[
\partial_t^2 T + \nabla \cdot (\partial_t q) = 0,
\]
and substituting $\partial_t q = -\tau^{-1} q - c^2 \nabla T$ leads to
\[
\partial_t^2 T - \tau^{-1} \nabla \cdot q - c^2 \nabla^2 T = 0.
\]
Using the continuity relation $\nabla \cdot q = -\partial_t T$, one obtains the telegrapher equation:
\begin{equation}
\tau \partial_t^2 T + \partial_t T = \alpha \nabla^2 T
\end{equation}
with $\alpha = c^2 \tau$, which interpolates between wave-like propagation at short times and diffusion at long times. In this extended-state formulation, the continuity relation $\partial_t T + \nabla \cdot q = 0$ serves as a kinematic closure constraint on the enlarged phase space $\psi = (T, q)$, ensuring energy conservation at the level of the dynamical system. Eq.~(4) describes finite-speed thermal propagation with characteristic velocity $c$, thereby removing the unphysical infinite propagation speed inherent in Fourier diffusion~\cite{ref22}. In the short-time regime $t \ll \tau$, the dynamics is wave-like and corresponds to a propagating second sound mode. In contrast, for $t \gg \tau$, the system relaxes to overdamped diffusive behavior. In the singular limit $\tau \to 0$, the flux equation reduces to the instantaneous constitutive relation $q = -\alpha \nabla T$, and the system collapses onto the classic diffusion equation:
\begin{equation}
\partial_t T = \alpha \nabla^2 T
\end{equation}
Thus, Fourier’s law emerges as a singular (fast-relaxation) limit of the underlying first-order dynamical system~\cite{ref23}, rather than as a fundamental law. At the microscopic level, heat transport is governed by the Boltzmann equation in the relaxation-time approximation~\cite{ref24}:
\begin{equation}
\partial_t f + \mathbf{v} \cdot \nabla f = -\tau^{-1}(f - f^{\mathrm{eq}})
\end{equation}
where $f$ is the distribution function and $f^{\mathrm{eq}}$ is the local equilibrium. Introducing macroscopic moments,
\[
T = \int f \, dv, \qquad q = \int \mathbf{v} f \, dv,
\]
and performing a first-order Chapman--Enskog truncation that retains only the lowest velocity moments~\cite{ref25}, one obtains the closed moment system
\[
\partial_t T + \nabla \cdot q = 0, \qquad
\tau \partial_t q + q = -\alpha \nabla T.
\]

This shows that the Cattaneo dynamics arises as the minimal representative hydrodynamic closure of kinetic theory~\cite{ref26}, establishing a direct correspondence between microscopic relaxation and finite-speed macroscopic transport. The resulting structure reveals a hierarchy of heat transport: kinetic (Boltzmann), mesoscopic (Cattaneo), and macroscopic (Fourier).

The system is naturally formulated in the Hilbert space
\[
\psi \in L^2(\mathbb{R}^d) \oplus L^2(\mathbb{R}^d;\mathbb{R}^d),
\]
with regularity $T \in H^1(\mathbb{R}^d)$ and $q \in H(\mathrm{div};\mathbb{R}^d)$. In this setting, the operator $\mathcal{V}$ is formally skew-adjoint, while $\Gamma$ is positive semi-definite, ensuring a well-posed dissipative evolution governed by a contraction semigroup.
\section{Energy dissipation and Lyapunov structure of the extended heat transport system}
We define the quadratic functional:
\begin{equation}
E(t) = \frac{1}{2} \int_{\mathbb{R}^d} \left( T^2 + \frac{1}{c^2} |q|^2 \right)\, dx
\end{equation}
which serves as a Lyapunov functional for the extended dynamical system~\cite{ref27}. It measures the combined contribution of the temperature field and the flux field in a quadratic norm, rather than the physical thermodynamic energy. This choice is not unique but represents the simplest diagonal metric rendering $\mathcal{V}$ skew‑adjoint and $\Gamma$ dissipative. We stress that this quadratic functional does not represent the thermodynamic internal energy, but rather a dynamical norm on the extended state space ensuring well-posedness and dissipative stability. Alternative admissible metrics related by bounded similarity transformations lead to equivalent generators with identical spectra, leaving the existence and order of the exceptional point invariant. Taking the time derivative yields:
\begin{equation}
\frac{dE}{dt} = \int_{\mathbb{R}^d} \left( T \, \partial_t T + \frac{1}{c^2} \, q \cdot \partial_t q \right)\, dx
\end{equation}
Substituting the evolution equations $\partial_t T = -\nabla \cdot q$ and $\partial_t q = -\tau^{-1} q - c^2 \nabla T$, we obtain
\begin{equation}
\frac{dE}{dt}
= \int_{\mathbb{R}^d}
\left(
- T \nabla \cdot q
- \frac{1}{c^2 \tau} |q|^2
- q \cdot \nabla T
\right)\, dx.
\tag{9}
\end{equation}

Assuming periodic boundary conditions or sufficient decay at infinity, integration by parts gives
\[
\int_{\mathbb{R}^d} (-T \nabla \cdot q)\, dx
=
\int_{\mathbb{R}^d} q \cdot \nabla T \, dx,
\]
so that the cross terms cancel exactly. This leads to the dissipation law
\begin{equation}
\frac{dE}{dt}
=
-\frac{1}{c^2 \tau}
\int_{\mathbb{R}^d}
|q|^2 \, dx
\le 0.
\tag{10}
\end{equation}

Thus, $E(t)$ is a Lyapunov functional for the dynamics, and its decay is governed entirely by the relaxation of the flux field. Dissipation vanishes only when $q = 0$ almost everywhere, corresponding to equilibrium states with vanishing flux. In the operator formulation $\partial_t \psi = -(\mathcal{V} + \Gamma)\psi$, the transport operator $\mathcal{V}$ is formally skew-adjoint and therefore conserves the quadratic functional, while the relaxation operator $\Gamma$ is positive semi-definite and solely responsible for irreversible decay.

The dissipation rate scales as $1/\tau$, indicating a crossover from weakly damped wave-like dynamics for large $\tau$ to strongly dissipative behavior in the small-$\tau$ regime. In the long-time limit, $q \to 0$ and $\nabla T \to 0$, and the system approaches a spatially uniform equilibrium state. In the singular limit $\tau \to 0$, the flux relaxes instantaneously according to
\[
q \approx -c^2 \tau \nabla T = -\alpha \nabla T.
\]
Substituting this into Eq.~(10) yields
\[
\frac{dE}{dt}
=
-\frac{\alpha}{c^2}
\int_{\mathbb{R}^d}
|\nabla T|^2 \, dx,
\]
which recovers the classic diffusive dissipation structure, where decay is governed by spatial gradients of the temperature field.

Defining the generalized entropy functional $S(t) = -E(t)$, one obtains $dS/dt \ge 0$, consistent with the second law of thermodynamics at the level of this reduced description. This structure shows that the system constitutes a damped hyperbolic dynamics in which diffusion emerges from flux relaxation, and irreversible dissipation is mediated entirely through the flux rather than directly through the temperature field. The formulation therefore provides a minimal representative first-order representation of heat transport with built-in dissipation and finite propagation speed.
\begin{figure}[htbp]
 \centering
        \includegraphics[width=1.0\textwidth]{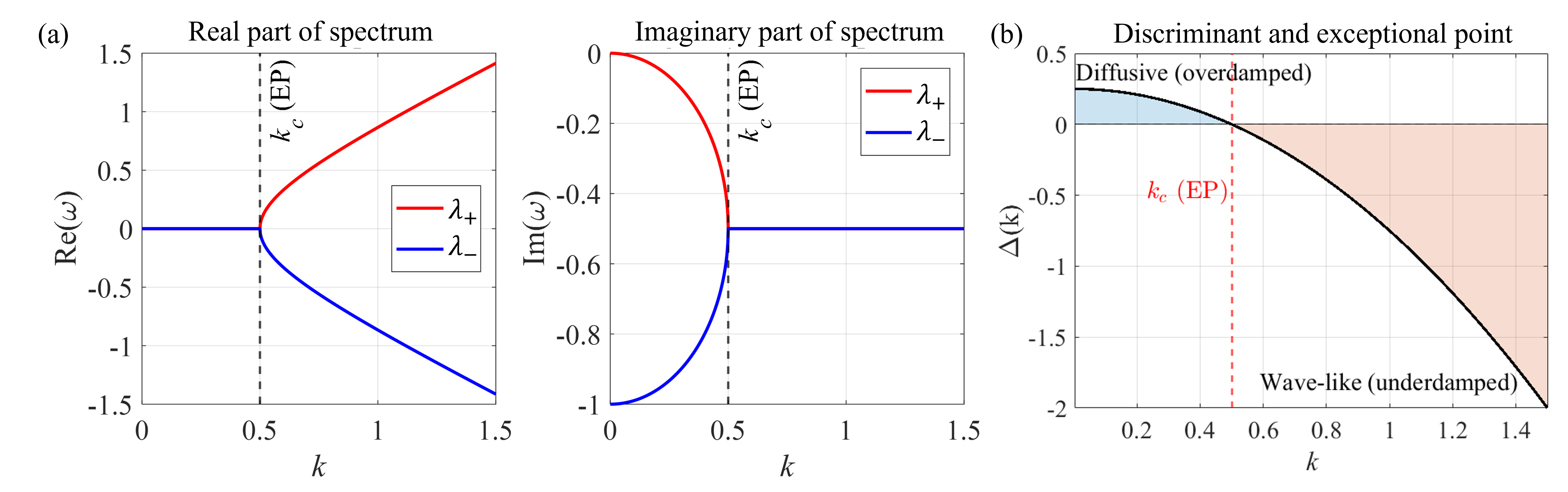}
 \caption{Non-Hermitian spectral structure of heat transport. (a) Dispersion relation of the eigenfrequencies as a function of wavevector $k$. The real part remains zero in the overdamped regime ($k < k_c$) and splits into propagating branches for $k > k_c$, indicating the emergence of propagating thermal modes. (b) Discriminant $\Delta(k)$ governing the spectral transition. The sign change of $\Delta$ separates diffusive ($\Delta > 0$) and wave-like ($\Delta < 0$) regimes. The exceptional point at $k_c$ marks the coalescence of eigenvalues and the onset of spectral bifurcation.}
\label{fig1}
\end{figure}
\section{Non-Hermitian spectral structure and dynamical regimes of the extended heat transport operator}
The thermal state vector
\[
\psi(x,t)=
\begin{pmatrix}
T(x,t) \\
q(x,t)
\end{pmatrix}
\in L^2(\mathbb{R}^d)\oplus L^2(\mathbb{R}^d;\mathbb{R}^d)
\]
evolves according to the first-order equation $\partial_t \psi = -\mathcal{L}\psi$, with operator decomposition $\mathcal{L} = \mathcal{V} + \Gamma$. Here, $\mathcal{V}$ represents the reversible transport coupling, while $\Gamma$ is positive semi-definite and defines a dissipative semigroup on the extended state space.

We consider plane-wave solutions of the form
\[
\psi(x,t) \sim e^{i k \cdot x + \lambda t} \, \hat{\psi},
\]
where $k$ is the wavevector and $\lambda \in \mathbb{C}$ is the temporal growth rate. This reduces the dynamics to the eigenvalue problem
\[
\lambda \hat{\psi} = -\mathcal{L}(k)\hat{\psi},
\]
with
\[
\mathcal{L}(k)=
\begin{pmatrix}
0 & i k \cdot \\
c^2 i k & \tau^{-1} I
\end{pmatrix}.
\]

This leads to the characteristic equation
\[
\lambda^2 + \tau^{-1}\lambda + c^2 k^2 = 0,
\]
which serves as the fundamental dispersion relation of the system.

To quantify the relative strength of transport and relaxation, we introduce the dimensionless Knudsen-type number
\[
\mathrm{Kn} = c k \tau.
\]

This parameter provides a natural measure of the competition between propagation and relaxation, organizing the spectrum into two asymptotic regimes:
(i) $\mathrm{Kn} \ll 1$, corresponding to the relaxation-dominated regime;
(ii) $\mathrm{Kn} \gg 1$, corresponding to the propagation-dominated regime.

We emphasize that the appearance of a critical wavenumber separating overdamped and underdamped thermal regimes in the Cattaneo or telegrapher equation is a well-established result. Likewise, the square-root structure of the spectrum at this transition is familiar from second-order linear dynamical systems. The purpose of the present formulation is not to rederive this transition, but to reinterpret it at the level of the non-Hermitian generator governing the extended temperature–flux dynamics.

The resulting spectral structure is summarized in Fig.~1. As shown in Fig.~1(a), the eigenfrequencies undergo a qualitative transition as a function of the wavevector $k$, separated by the critical scale $k_c$. In the long-wavelength regime ($k<k_c$), both eigenvalues are purely real, corresponding to overdamped diffusive relaxation. Beyond the critical point ($k>k_c$), the spectrum bifurcates into a complex-conjugate pair, signaling the onset of propagating thermal modes.

This transition is governed by the discriminant $\Delta(k)$, shown in Fig.~1(b), whose sign determines the dynamical regime. The vanishing of $\Delta$ at $k=k_c$ marks an exceptional point (EP), where both eigenvalues and eigenvectors coalesce.

The eigenvalues are
\begin{equation}
\lambda_{\pm}(k)
=
-\frac{1}{2\tau}
\pm
\sqrt{\frac{1}{4\tau^2} - c^2 k^2}.
\tag{11}
\end{equation}

A critical wavenumber $k_c = (2c\tau)^{-1}$ separates overdamped and underdamped regimes. For $|k| \ll k_c$, expansion yields
\begin{equation}
\sqrt{\frac{1}{4\tau^2} - c^2 k^2}
=
\frac{1}{2\tau}
\left(1 - 2c^2 \tau^2 k^2 + \mathcal{O}(k^4)\right).
\tag{12}
\end{equation}

Thus,
\[
\lambda_+(k) = -\alpha k^2 + \mathcal{O}(\tau k^4), 
\qquad
\lambda_-(k) = -\tau^{-1} + \mathcal{O}(k^2),
\]
with $\alpha = c^2 \tau$. Here, $\lambda_+$ defines a slow hydrodynamic manifold~\cite{ref28}, while $\lambda_-$ corresponds to a rapidly decaying non-hydrodynamic flux mode.

For $|k| \gg k_c$,
\begin{equation}
\lambda_{\pm}(k)
=
-\frac{1}{2\tau}
\pm i c |k|
+ \mathcal{O}\!\left(\frac{1}{\tau k^2}\right).
\tag{13}
\end{equation}

This regime corresponds to damped propagating thermal waves (second sound), where propagation and dissipation coexist. The spectrum naturally splits as
\begin{equation}
\sigma(\mathcal{L}(k)) = \sigma_{\mathrm{diff}}(k) \cup \sigma_{\mathrm{wave}}(k),
\tag{14}
\end{equation}
with
\[
\sigma_{\mathrm{diff}}(k) = \{-\alpha k^2 + \mathcal{O}(\tau k^4)\}, 
\qquad
\sigma_{\mathrm{wave}}(k) = \left\{-\frac{1}{2\tau} \pm i c |k| \right\}.
\]

Because $\mathcal{L}(k)$ is non-normal ($\mathcal{L}\mathcal{L}^\dagger \neq \mathcal{L}^\dagger \mathcal{L}$), its eigenvectors form a biorthogonal system:
\begin{equation}
\mathcal{L}(k) r_\pm = -\lambda_\pm r_\pm, 
\qquad
\mathcal{L}^\dagger(k) l_\pm = -\overline{\lambda_\pm} l_\pm,
\tag{15}
\end{equation}
with normalization $\langle l_i, r_j \rangle = \delta_{ij}$.

This yields the resolution of identity
\[
I = \sum_{n=\pm} |r_n\rangle \langle l_n|.
\]

The field admits the expansion
\begin{equation}
\psi(k,t) = \sum_{n=\pm} r_n(k) a_n(k,t),
\qquad
a_n = \langle l_n, \psi \rangle,
\tag{16}
\end{equation}
and each mode evolves independently:
\[
\partial_t a_n(k,t) = -\lambda_n(k) a_n(k,t).
\]

Thus, diagonalization occurs only in the biorthogonal basis, rather than in the physical field representation. At $k = k_c$, $\lambda_+(k_c) = \lambda_-(k_c) = -1/(2\tau)$, the algebraic multiplicity exceeds the geometric multiplicity, and the operator becomes defective.
\begin{figure}[htbp]
 \centering
        \includegraphics[width=0.6\textwidth]{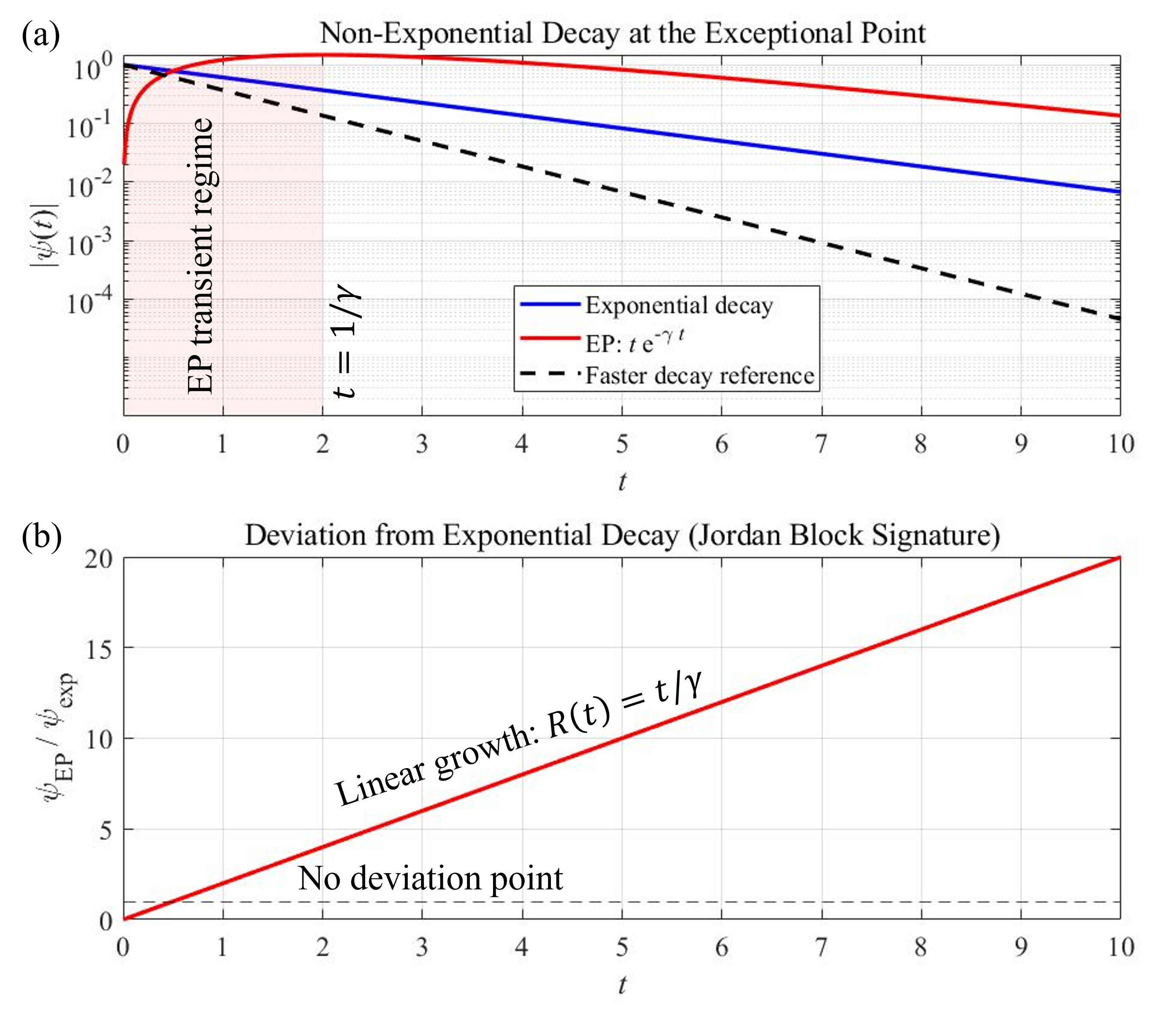}
 \caption{Breakdown of exponential relaxation at an exceptional point (EP). (a) Time evolution of the field amplitude $|\psi(t)|$ is shown on a logarithmic scale for three cases: standard exponential decay $e^{-\gamma t}$, EP dynamics $t e^{-\gamma t}$, and a faster reference decay $e^{-2\gamma t}$. The EP case exhibits a polynomial prefactor arising from the non-diagonalizable (Jordan block) structure of the underlying generator, leading to a deviation from pure exponential relaxation. (b) Linear-in-time growth characteristic of EP-induced non-exponential dynamics, shown via the ratio $\psi_{\mathrm{EP}}/\psi_{\mathrm{exp}} = t/\gamma$. The dashed horizontal line indicates the reference level $R(t)=1$, used to highlight the onset of deviation from the standard exponential envelope.}
\label{fig2}
\end{figure}

In this case,
\[
\mathcal{L}(k_c) = \lambda_c I + N, \qquad N^2 = 0,
\]
leading to
\[
e^{-\mathcal{L}(k_c)t} = e^{-\lambda_c t}(I - Nt),
\]
which gives rise to non-exponential transient dynamics with a polynomial prefactor~\cite{ref29}. A direct consequence of eigenmodes coalescence at the exceptional point. This behavior is illustrated in Fig.~2. As shown in Fig.~2(a), the EP dynamics exhibits a clear deviation from pure exponential decay. The ratio shown in Fig.~2(b) highlights the emergence of linear-in-time growth, which is a hallmark of the underlying Jordan block structure. Away from the EP, the operator becomes diagonalizable and standard modal decomposition is recovered

In the long-wavelength limit, eliminating the fast flux mode yields $q \approx -c^2 \tau \nabla T = -\alpha \nabla T$, recovering Fourier diffusion $\partial_t T = \alpha \nabla^2 T$. This shows that diffusion corresponds to a collapsed spectral manifold of a non-Hermitian hyperbolic operator~\cite{ref30}. Importantly, this behavior cannot be mimicked by a generic superposition of underdamped modes with distinct frequencies. In a diagonalizable system, the temporal response remains expressible as a sum of exponentials and oscillatory terms, whereas the EP-induced dynamics requires a non-normal generator with coalescing eigenvectors. The presence of polynomial prefactors therefore provides an unambiguous criterion to distinguish exceptional-point dynamics from ordinary damped wave propagation. We stress that the polynomial-in-time prefactor arising at the exceptional point is not merely a mathematical curiosity, but provides a clear dynamical signature distinguishing EP dynamics from a conventional underdamped response. In contrast to ordinary damped oscillations, which exhibit purely exponential envelopes, the EP response displays a transient algebraic enhancement that can be isolated by rescaling out the exponential decay. This feature is directly accessible in time-resolved measurements following a pulse excitation and does not require precise phase coherence between modes. While the overdamped-underdamped transition is often discussed solely in terms of eigenvalues, the exceptional point entails an additional structural consequence: the coalescence of eigenvectors and the failure of modal diagonalization. As a result, the temporal evolution acquires polynomial prefactors that cannot be captured within a conventional normal-mode analysis.

Higher-order gradient corrections, non-local memory kernels, or additional dissipative couplings can in principle be incorporated within the same operator framework. These extensions promote the dispersion relation to higher-order or non-polynomial spectral equations. Importantly, as long as the extended dynamics retains a finite-dimensional slow manifold coupled to a fast relaxing flux variable, the lowest-order branch-point singularity persists. In this sense, the exceptional point identified here is structurally stable: additional terms deform, but do not eliminate, the diffusion-wave transition encoded in the coalescence of spectral branches.

\section{Non-Hermitian exceptional point and spectral transition in heat transport}
The spectral structure in Eq.~(1) exhibits a transition governed by the competition between relaxation and propagation encoded in the discriminant
\begin{equation}
\Delta(k) = \frac{1}{4\tau^2} - c^2 k^2.
\tag{17}
\end{equation}

The vanishing of $\Delta(k)$ defines a critical wavenumber $k_c = (2c\tau)^{-1}$, at which the two eigenvalue branches coalesce into a single degenerate eigenvalue
\[
\lambda_+(k_c) = \lambda_-(k_c) = -\frac{1}{2\tau}.
\]

To characterize the behavior near the transition, we set $k = k_c + \delta k$. Expanding the discriminant yields
\begin{equation}
\Delta(k) \approx -2c^2 k_c \, \delta k.
\tag{18}
\end{equation}

This leads to the local form of the eigenvalues
\[
\lambda_{\pm}(k)
=
-\frac{1}{2\tau}
\pm
\sqrt{-2c^2 k_c \, \delta k},
\]
which exhibits a square-root branch point at $k = k_c$. As a result, the spectrum is locally two-sheeted and non-analytic at the transition.

The evolution of the complex spectrum across this transition is visualized in Fig.~3. The eigenvalues trace continuous trajectories in the complex plane as $k$ increases, coalescing at the exceptional point (EP) and subsequently bifurcating into distinct branches. The arrows indicate the direction of spectral flow, revealing the underlying branch-point topology. This behavior directly characterizes the diffusion–wave transition in spectral space.

\begin{figure}[htbp]
 \centering
        \includegraphics[width=0.6\textwidth]{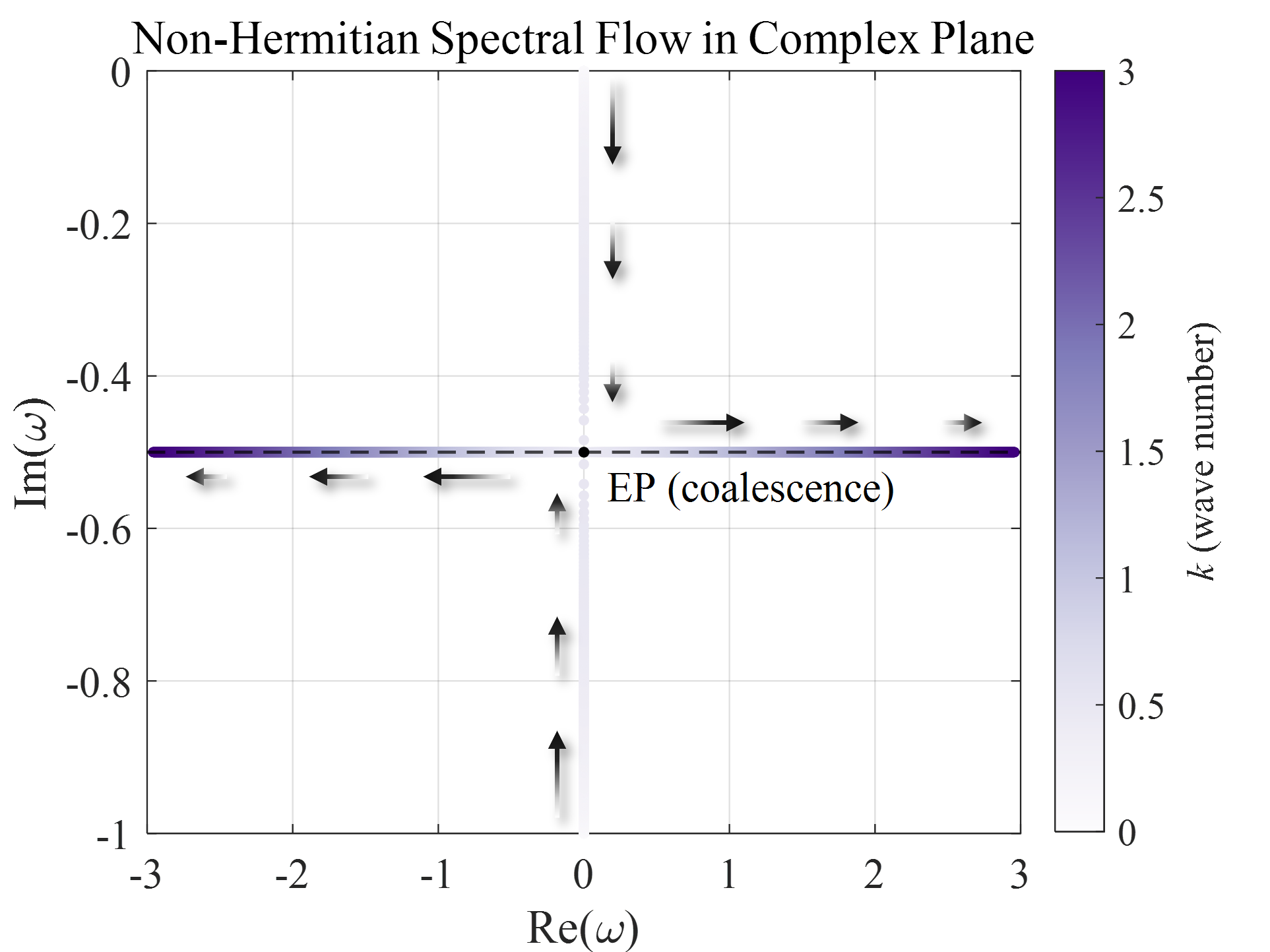}
 \caption{Complex eigenvalue spectrum in the non-Hermitian plane. The color gradient indicates increasing wavevector k, while arrows denote the direction of spectral evolution. The eigenvalues coalesce at the EP and bifurcate into distinct branches beyond it, revealing the nonanalytic structure underlying the diffusion-wave transition.}
\label{fig3}
\end{figure}
The sign of $\Delta(k)$ determines the local spectral character. For $\Delta(k) > 0$, the eigenvalues are purely real, corresponding to monotonic relaxation. For $\Delta(k) < 0$, the spectrum forms a complex-conjugate pair, leading to oscillatory decay. Thus, the transition is not a smooth crossover but a non-analytic change in spectral topology.

At the critical wavenumber $k = k_c$, the spectral degeneracy leads to a defective generator~\cite{ref31}. In this regime, the eigenbasis becomes incomplete and the evolution cannot be fully represented by modal decomposition. Instead, the operator admits a local Jordan structure~\cite{ref32}
\[
\mathcal{L}(k_c) = \lambda_c I + N, \qquad N^2 = 0,
\]
which modifies the temporal evolution beyond exponential decay. As a consequence, the propagator acquires a non-modal correction:
\[
e^{-\mathcal{L}(k_c)t} = e^{-\lambda_c t}(I - Nt).
\]
This polynomial prefactor reflects the coalescence of eigenmodes and leads to transient dynamics that are not captured by standard eigenfunction expansion.

For $k \neq k_c$, the Green's function is obtained from the spectral representation
\begin{equation}
G(x,t) =
\int e^{i k \cdot x}
\sum_{\pm}
e^{-\lambda_{\pm}(k)t}
\, dk.
\tag{19}
\end{equation}

In the long-time limit, the spatial response separates into contributions from distinct spectral sectors:
\[
G(x,t) \sim G_d(x,t) + G_b(x,t) + G_{\mathrm{EP}}(x,t),
\]
where the diffusive contribution is
\[
G_d(x,t) \sim t^{-1/2} \exp\!\left(-\frac{x^2}{4\alpha t}\right),
\]
together with propagating and exceptional-point contributions $G_b(x,t)$ and $G_{\mathrm{EP}}(x,t)$.

These terms should be interpreted as asymptotic components arising from different regions of spectral integration, rather than independent dynamical modes.

To characterize the propagating component $G_b(x,t)$, we consider an initial Fourier-space Gaussian wave packet centered at $k_0 > k_c$:
\begin{equation}
\hat{\psi}(k,0) = A \exp\!\left[-\frac{(k-k_0)^2}{2\sigma^2}\right].
\tag{21}
\end{equation}

The time evolution in the oscillatory regime is governed by
\[
\hat{\psi}(k,t) = \hat{\psi}(k,0)\, e^{-\lambda_+(k)t},
\]
with
\[
\lambda_+(k) = -\frac{1}{2\tau} + i \omega(k),
\qquad
\omega(k) = \sqrt{c^2 k^2 - \frac{1}{4\tau^2}}.
\]

Expanding $\lambda_+(k)$ around $k_0$ yields
\begin{equation}
\lambda_+(k)
\approx
\lambda_+(k_0)
+ (k-k_0)\lambda_+'(k_0)
+ \frac{1}{2}(k-k_0)^2 \lambda_+''(k_0).
\tag{22}
\end{equation}

Applying the stationary phase approximation~\cite{ref33}, we obtain the real-space asymptotic form
\begin{equation}
\psi(x,t)
\sim
e^{-t/2\tau}
\exp\!\left[-\frac{(x - v_g t)^2}{2\sigma_x^2(t)}\right]
e^{i(k_0 x - \omega_0 t)},
\tag{23}
\end{equation}
where the group velocity is
\[
v_g = \frac{d\omega}{dk}
=
\frac{c^2 k_0}{\sqrt{c^2 k_0^2 - \frac{1}{4\tau^2}}}.
\]

This velocity satisfies $v_g \to c$ for $k_0 \gg k_c$, while $v_g \to \infty$ as $k_0 \to k_c^+$, indicating a critical anomaly. This divergence does not imply superluminal transport, but rather signals the breakdown of the single-wave-packet (quasiparticle-like) approximation due to eigenmode coalescence.

The second-order expansion yields an effective spreading
\[
\sigma_x^2(t) \sim \sigma^{-2} + D_{\mathrm{eff}} t,
\]
indicating that propagation and dispersion remain intrinsically coupled, reflecting the non-normal nature of the generator. As $k_0 \to k_c$, both $v_g$ and $D_{\mathrm{eff}}$ become singular.

In this regime, the eigenmode expansion ceases to be uniformly valid, and the dynamics is dominated by the Jordan contribution:
\[
\psi(x,t) \sim t\, e^{-t/2\tau},
\]
revealing non-modal transient amplification near the exceptional point (EP).
\begin{figure}[htbp]
 \centering
        \includegraphics[width=1.0\textwidth]{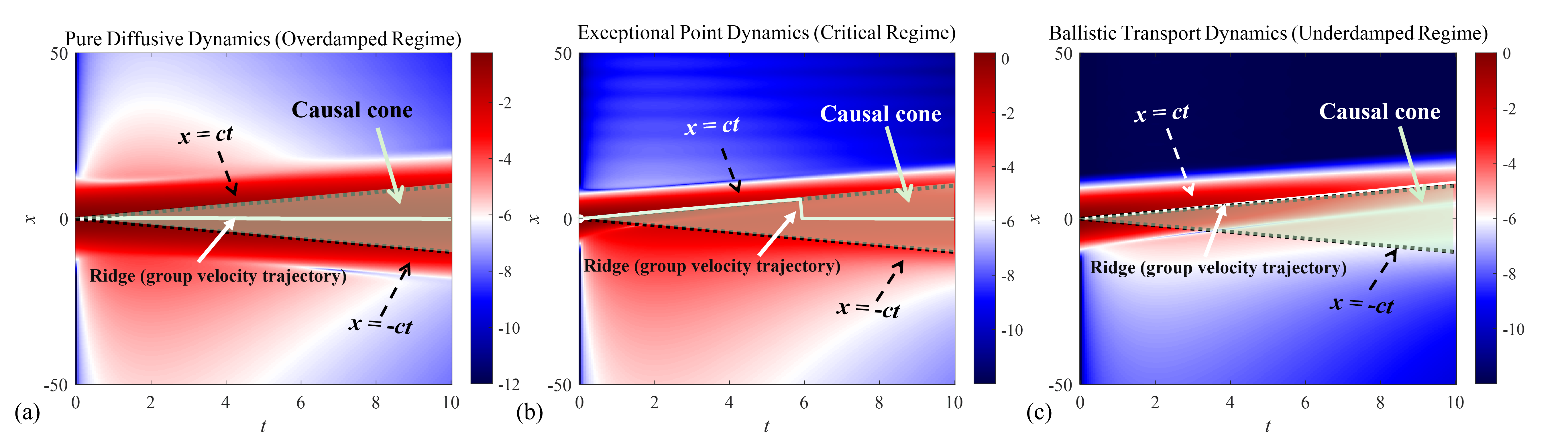}
 \caption{Dynamical regimes across the exceptional point in non-Hermitian heat transport. (a) Overdamped regime ($k < k_c$): The wave packet exhibits purely diffusive behavior, characterized by a stationary ridge at $x = 0$ and monotonic spatial broadening without ballistic transport. (b) Near the exceptional point ($k \approx k_c$): The dynamics become nontrivial, featuring pronounced deformation of the wave packet. The ridge bends due to the reshaping of the dispersion relation, while interference fringes emerge from non-Hermitian phase mixing. A diffuse background persists due to overdamped modes. (c) Underdamped regime ($k > k_c$): Ballistic propagation dominates, forming a clear ridge corresponding to a well-defined group velocity. The wave packet maintains a coherent structure with reduced distortion, and the dynamics are confined within the causal cone $x = \pm ct$.}
\label{fig4}
\end{figure}
The resulting dynamical regimes are illustrated in Fig.~4. In the overdamped regime (Fig.~4(a)), the wave packet remains localized and exhibits purely diffusive spreading without ballistic transport. Near the exceptional point (Fig.~4(b)), the dynamics becomes strongly nontrivial, with pronounced distortion and enhanced interference due to non-Hermitian phase mixing. In the underdamped regime (Fig.~4(c)), a well-defined propagating front emerges, characterized by finite group velocity and reduced distortion.

These results establish a direct correspondence between spectral topology and real-space dynamics, showing that the EP governs not only the spectral structure but also the qualitative nature of thermal transport.

To quantify the onset of oscillatory behavior, we define $\phi(k) = \mathrm{Im}\,\lambda_+(k)$. Near $k_c$, using Eq.~(18),
\begin{equation}
\phi(k)
=
\begin{cases}
0, & k < k_c, \\
\sqrt{2c^2 k_c (k - k_c)}, & k > k_c,
\end{cases}
\tag{24}
\end{equation}

This square-root onset defines a critical exponent $\beta = 1/2$, indicating a branch-point singularity rather than a smooth crossover. The transition corresponds to the emergence of oscillatory thermal modes, i.e., the onset of underdamped heat propagation.

For $k < k_c$, the system exhibits monotonic relaxation, while for $k > k_c$, damped propagating modes appear. This behavior should be interpreted as a change in the spectral nature of thermal response functions, rather than a thermodynamic phase transition~\cite{ref34}.

From a dynamical systems perspective, the transition separates a regime of diffusive spreading with unbounded support and a regime of finite-velocity propagation with wavefront structure. Thus, the spectral transition governs the qualitative nature of thermal information transport.

\section{Non-Hermitian spectral topology of heat transport}
The spectral structure defined by Eq.~(1) can be written as the algebraic relation
\[
D(\lambda,k) = \lambda^2 + \tau^{-1}\lambda + c^2 k^2,
\]
which defines a complex spectral curve in the $(\lambda,k)$ plane. The corresponding eigenvalues
\[
\lambda_{\pm}(k) = -\frac{1}{2\tau} \pm \sqrt{\frac{1}{4\tau^2} - c^2 k^2}
\]
are naturally interpreted as branches of a multi-valued analytic function~\cite{ref35}. Due to the square-root dependence, $\lambda(k)$ is defined on a two-sheeted Riemann surface over complex $k$-space.

The two sheets correspond to distinct spectral sectors (overdamped and underdamped), which are analytically connected through branch points determined by $\Delta(k)=0$. Near the critical point $k_c$, introducing $z = k - k_c$ yields
\[
\lambda_{\pm}(k) = -\frac{1}{2\tau} \pm \sqrt{-2c^2 k_c z},
\]
explicitly exhibiting a square-root branch-point singularity.
\begin{figure}[htbp]
 \centering
        \includegraphics[width=0.6\textwidth]{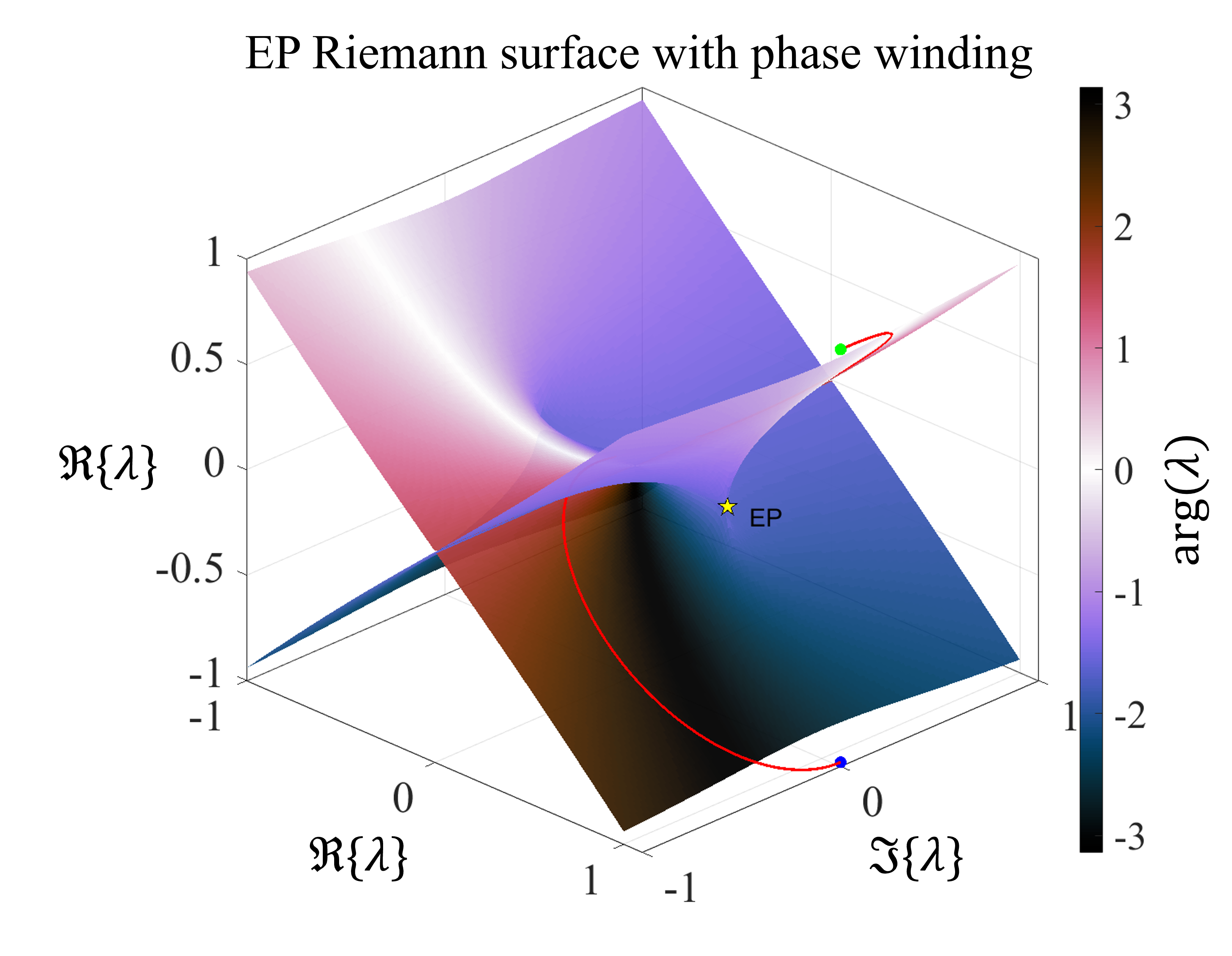}
 \caption{Three-dimensional Riemann surface of the non-Hermitian spectrum with phase winding and exceptional-point (EP) topology. The real part of the complex eigenfrequency $\omega(k)$ is plotted over the complex momentum plane $k = k_x + i k_y$, forming a two-sheeted Riemann surface corresponding to the $\omega_{\pm}$ branches. The color encodes the phase $\arg(\omega)$, revealing a nontrivial phase-winding structure around the EP, where the two sheets coalesce. A closed loop in complex $k$-space encircling the EP is projected onto the spectral surface, showing permutation of eigenvalues after a single encircling and restoration after a double encircling, consistent with square-root branch-point topology. This behavior directly visualizes the non-Hermitian holonomy associated with EP-induced spectral branching.}
\label{fig5}
\end{figure}
This multi-valued spectral structure is visualized in Fig.~5 as a two-sheeted Riemann surface. The exceptional point acts as a branch point connecting the two sheets. A closed loop encircling the EP,
\[
k(\theta) = k_c + \rho e^{i\theta}, \qquad \theta \in [0,2\pi],
\]
induces a permutation of the spectral branches $\lambda_+(k) \leftrightarrow \lambda_-(k)$, while a second encircling restores the original branch. This defines a non-trivial monodromy, showing that $\lambda(k)$ cannot be defined globally as a single-valued function.

The branch structure can be quantified through the winding of the spectral gap function:
\begin{equation}
W = \frac{1}{2\pi i} \oint_C \frac{d}{dk} \log\!\big(\lambda_+(k) - \lambda_-(k)\big)\, dk,
\tag{25}
\end{equation}
where $C$ is a closed contour enclosing the EP. Using $\lambda_+(k)-\lambda_-(k)=2\sqrt{\Delta(k)}$, one obtains a half-integer winding $W=1/2$, reflecting the square-root nature of the singularity. This half-integer value is a hallmark of branch-point topology in non-Hermitian systems.

The nontrivial analytic structure of $\lambda(k)$ has direct dynamical consequences. Because the eigenvalue branches are not globally separable, spectral decomposition becomes non-uniform near the branch point, leading to a breakdown of purely exponential modal relaxation. As a result, temporal responses acquire corrections beyond simple exponential decay, reflecting the underlying non-normal and multi-sheeted spectral geometry.

This branch-point topology encodes the transition between qualitatively distinct dynamical regimes: purely relaxational dynamics and oscillatory propagation. It therefore provides a unified geometric framework for the diffusion–wave transition in heat transport.

Importantly, this topology does not introduce a new dispersion relation, but rather provides a geometric and spectral organization principle that unifies diffusive relaxation, wave propagation, and non-modal transient dynamics within a single analytic framework.

\section{Anisotropic extension and directional heat transport}
We extend the first-order formulation to anisotropic media~\cite{ref36}, where transport explicitly depends on spatial direction. In this case, the physically constrained linear local operator, consistent with the extended state vector $\psi = (T, q_i)$, takes the block form
\begin{equation}
L =
\begin{pmatrix}
0 & \partial_i \\
C_{ij}\partial_j & \Gamma_{ij}
\end{pmatrix},
\tag{26}
\end{equation}
where $C_{ij}$ is a symmetric positive-definite tensor encoding direction-dependent propagation, and $\Gamma_{ij}$ is a positive semi-definite relaxation operator.

Physically, such a tensorial relaxation structure naturally arises in crystalline or layered media, where phonon scattering rates depend on both polarization and propagation direction. In anisotropic crystals such as graphite, hexagonal boron nitride, and van der Waals heterostructures, longitudinal and transverse phonon branches exhibit distinct relaxation times along different crystallographic axes. In this setting, $\Gamma_{ij}$ can be interpreted as an effective hydrodynamic relaxation tensor obtained by projecting the microscopic phonon collision operator onto the lowest-order moments.

The resulting evolution equations read
\begin{equation}
\begin{cases}
\partial_t T + \partial_i q_i = 0, \\
\partial_t q_j + (\Gamma q)_j = - C_{ij} \partial_i T,
\end{cases}
\tag{27}
\end{equation}
which generalize the isotropic Cattaneo system to anisotropic transport. Eliminating the flux field $q_i$ yields a second-order anisotropic damped wave equation for the temperature field,
\begin{equation}
\partial_t^2 T + \partial_i (\Gamma q)_i = C_{ij}\partial_i \partial_j T.
\tag{28}
\end{equation}

In the isotropic limit $\Gamma = \tau^{-1} I$, Eq.~(28) reduces to the standard telegrapher equation. In the general anisotropic case, however, the relaxation term retains its full tensorial structure and cannot be reduced to a scalar damping rate.

Assuming plane-wave solutions of the form $T \sim e^{i(\mathbf{k}\cdot\mathbf{x}-\omega t)}$, we obtain the dispersion relation
\begin{equation}
\omega^2 + i\omega \left( \hat{k}_i \Gamma_{ij} \hat{k}_j \right)
= C_{ij} k_i k_j,
\tag{29}
\end{equation}
where $\hat{\mathbf{k}} = \mathbf{k}/|\mathbf{k}|$. The impact of anisotropy is illustrated in Fig.~6. The isofrequency contours are deformed from circular to elliptical, reflecting direction-dependent propagation properties~\cite{ref37}. The red contour marks the exceptional surface separating overdamped and propagating regimes. This demonstrates that the exceptional point generalizes to a direction-dependent manifold~\cite{ref38}, enabling intrinsic steering of heat flow.

The directional effects discussed here concern energy transport rather than phase propagation. In the hyperbolic regime, the heat flux becomes dynamically slaved to the propagating mode and aligns with the group velocity, which governs energy transport direction. The resulting steering effect therefore corresponds to a genuine redirection of energy flow, rather than a phase or interference phenomenon. This distinction is particularly important in anisotropic media, where group-velocity surfaces generally deviate from spherical symmetry.

Defining the directional propagation scale
\[
\Omega^2(\mathbf{k}) = C_{ij}k_i k_j,
\]
the spectrum depends explicitly on the propagation direction. The transition between overdamped and underdamped regimes is determined by the condition
\[
\Omega^2(\mathbf{k}) = \frac{1}{4}
\left( \hat{k}_i \Gamma_{ij} \hat{k}_j \right)^2.
\]
This defines a direction-dependent critical surface in momentum space, replacing the single critical point of the isotropic case. Consequently, different propagation directions enter the wave-like regime at different thresholds.

The group velocity follows from the dispersion relation,
\begin{equation}
v_g^i = \frac{\partial \omega}{\partial k_i}
= \frac{C_{ij}k_j}{\omega}.
\tag{30}
\end{equation}

In general, this implies $\mathbf{v}_g \nparallel \mathbf{k}$. Since the heat flux is dynamically linked to the propagating mode, $q \sim \mathbf{v}_g$, this leads to $q \nparallel \nabla T$, in contrast to Fourier’s law. This non-collinearity represents a breakdown of isotropic transport assumptions: energy propagation is no longer aligned with wavefront normals, and transport becomes intrinsically tensorial rather than scalar. As a result, anisotropy acts as a thermal steering mechanism, enabling directional control of heat flow through engineering of $C_{ij}$.

We emphasize that this non-collinearity does not violate Onsager reciprocity. The linear response relations remain symmetric at the microscopic level; the apparent misalignment arises from anisotropic dispersion and wave propagation effects rather than antisymmetric transport coefficients. Similar behavior is well known in anisotropic wave systems, including phononic and elastic media, where energy flux (group velocity) is generally not parallel to the wavevector.

Finally, the discriminant of Eq.~(29) generalizes to
\begin{equation}
\Delta(\mathbf{k}) =
\frac{1}{4} \left( \tau^{-1}_{\mathrm{eff}} \right)^2
- C_{ij}k_i k_j.
\tag{31}
\end{equation}

The degeneracy condition $\Delta(\mathbf{k})=0$ defines a direction-dependent exceptional surface in momentum space, rather than an isolated point. This surface encodes the competition between anisotropic propagation and relaxation mechanisms.

As a concrete example, layered crystals such as graphite and black phosphorus provide natural platforms where strong in-plane versus out-of-plane transport anisotropy is well established. In these systems, the interplay between direction-dependent relaxation and propagation leads to anisotropic dispersion surfaces, suggesting that the exceptional surfaces described here could, in principle, be probed via time-resolved thermal pulse experiments or numerical transport simulations.

The dynamical signatures discussed here are directly relevant to systems exhibiting second-sound or non-Fourier heat transport, such as low-temperature crystalline solids and semiconductors where Cattaneo-type dynamics has been experimentally established. In such systems, time-resolved thermal grating or pump–probe techniques provide access to transient temperature and heat-flux dynamics on the relevant relaxation timescales. At the nanoscale, non-diffusive heat transport has been extensively investigated using molecular dynamics and kinetic simulations. The extended temperature–flux framework employed here can be directly implemented in numerical schemes, enabling controlled comparison between EP-dominated transient dynamics and conventional underdamped propagation in finite or extended geometries. In disordered or glassy systems, where relaxation and propagation compete on comparable timescales, the non-normal nature of the transport generator becomes particularly pronounced. While strong disorder may smear sharp spectral features in frequency space, the associated non-modal transient dynamics in the time domain remain a robust indicator of proximity to an exceptional point.

We finally comment on the role of noise, nonlinearities, and boundary effects. Weak noise and moderate nonlinearities perturb but do not destroy the defective structure of the generator, leading instead to a smooth crossover rather than an abrupt disappearance of EP-related features. Finite boundaries discretize the spectrum but preserve mode coalescence near the critical surface. As a result, the non-modal transient response associated with the exceptional surface is expected to remain observable over a finite parameter window, rather than being a singular artifact of an idealized continuum model.

\section{Conclusion}
In this work, we have introduced a minimal representative first-order non-Hermitian operator framework that unifies ballistic and diffusive heat transport within a single dynamical description. By elevating temperature and heat flux to extended state vector, the resulting evolution operator naturally incorporates conservation, dissipation, and finite propagation speed, thereby reconciling Fourier diffusion and Cattaneo-type transport as different spectral regimes of the same underlying generator. From the operator-theoretic perspective, this transition corresponds to an exceptional point of the non-Hermitian generator, where both eigenvalues and eigenvectors coalesce. This identification goes beyond the standard overdamped-underdamped classification by highlighting the defective nature of the generator and the ensuing non-diagonalizable dynamics. This transition is governed by a branch-point singularity rather than a smooth crossover, resulting in mode coalescence, nonanalytic dispersion, and nonmodal transient dynamics. Within this picture, Fourier’s law emerges as a singular fast-relaxation limit corresponding to the collapse of a spectral manifold, instead of a fundamental transport principle. The framework admits a natural extension to anisotropic media, where the exceptional point generalizes to direction-dependent exceptional surfaces, enabling intrinsically anisotropic and non-collinear heat flow. These results identify non-Hermitian spectral topology as a unifying organizing principle for thermal dynamics across scales. Beyond heat transport, the present formulation establishes a general paradigm for dissipative transport phenomena in which hyperbolic propagation, diffusion, and irreversible relaxation coexist within a unified non-Hermitian dynamical structure.

%
% Each of the commands below will create an unnumbered section with the appropriate heading.
% Remove any sections that are not relevant for your article.
% All sections except suppdata will be removed if the [anonymous] option is used.
% See iopjournal-guidelines.pdf for more information.
%

\ack{P. Z. acknowledges the support from the Adolf Martens Postdoctoral Fellowship (BAM-AMF-2025-1).}

\data{All data that support the findings of this study are included within the article (and any supplementary files).}
% For more information on IOP Publishing's research data policy see: https://publishingsupport.iopscience.iop.org/questions/research-data/

\appendix

\renewcommand{\thesection}{Appendix \Alph{section}}

\section{Minimality of the first-order heat transport operator}
The proposed formulation can be interpreted as a first-order coupled field theory with Dirac-type structure,
\begin{equation}
(i\Gamma^\mu \partial_\mu - M)\psi = 0, \tag{A1}
\end{equation}
where $\psi = (T,q)$ combines temperature and heat flux into an extended state vector. Assuming $\Gamma^0$ is invertible, this equation can be recast in evolutionary form $\partial_t \psi = -L\psi$.

We emphasize that by “minimal first-order operator” we do not claim uniqueness in a fully general functional space. Instead, minimality is understood in a restricted and physically motivated sense. Specifically, we consider the class of linear, local, first-order extensions of the energy balance equation that: (i) preserve exact conservation at the level of the extended state, (ii) ensure finite signal propagation speed, (iii) respect isotropy, and (iv) admit a coercive Lyapunov functional generating a contraction semigroup. Within this constrained class, the resulting operators form a finite-dimensional equivalence family up to rescaling and similarity transformations. The operator introduced here should therefore be regarded as a canonical representative rather than a unique realization.

The evolution equation can be written as
\begin{equation}
\partial_t \psi = -\mathcal{L}\psi, \qquad
\psi \in \mathcal{L}^2(\mathbb{R}^d) \oplus \mathcal{L}^2(\mathbb{R}^d;\mathbb{R}^d),
\tag{A2}
\end{equation}
where $\mathcal{L}$ is a differential operator of at most first order in space. Under isotropy, the physically constrained block structure takes the form
\begin{equation}
\mathcal{L} =
\begin{pmatrix}
A & B^i \partial_i \\
C^i \partial_i & D
\end{pmatrix},
\tag{A3}
\end{equation}
where $A$ and $D$ are scalar and vector-valued operators respectively, and $B^i, C^i$ are tensors constrained by isotropy.

We impose the following physically motivated conditions: (i) In the absence of sources, temperature satisfies a continuity equation $\partial_t T + \nabla \cdot q = 0$, which implies $A=0$ and $B^i \partial_i q = \nabla \cdot q$. (ii) To avoid instantaneous (parabolic) transport, the flux dynamics must include a first-order coupling to the temperature gradient, $\partial_t q \sim -\nabla T$, excluding higher-order spatial derivatives and restricting admissible couplings to first-order operators, yielding $C^i \partial_i T = c^2 \nabla T$. (iii) For isotropic media, tensorial couplings reduce to scalar multiples of the identity, ensuring rotational invariance. (iv) In the absence of gradients, the flux relaxes toward equilibrium, $\partial_t q = -\Gamma q$, where $\Gamma$ is a positive semi-definite operator. (v) We require the existence of a quadratic Lyapunov functional ensuring dissipative contraction dynamics.
\begin{equation}
E[\psi] = \frac{1}{2} \int \big( T^2 + \beta |q|^2 \big)\, dx,
\tag{A4}
\end{equation}
such that $\frac{dE}{dt} \le 0$. This condition constrains $\Gamma$ to be symmetric positive semi-definite and excludes non-dissipative contributions in the relaxation term. Under the above assumptions, the generator reduces (up to rescaling of units) to
\begin{equation}
\mathcal{L} =
\begin{pmatrix}
0 & \nabla \cdot \\
c^2 \nabla & \Gamma
\end{pmatrix}.
\tag{A5}
\end{equation}

In the simplest case $\Gamma = \tau^{-1} I$, this yields the minimal representative form
\begin{equation}
\mathcal{L}_{\mathrm{min}} =
\begin{pmatrix}
0 & \nabla \cdot \\
c^2 \nabla & \tau^{-1} I
\end{pmatrix}.
\tag{A6}
\end{equation}

This construction shows that the proposed system is not arbitrary, but represents the minimal first-order closure consistent with: (i) conservation, (ii) finite propagation speed, (iii) isotropy, and (iv) dissipative stability. Higher-order spatial derivatives or additional couplings correspond to non-minimal extensions beyond this leading-order structure.

\section{Singular perturbation structure of the Fourier limit}
We consider the isotropic first-order dynamics in Eq.~(3), where $\tau > 0$ denotes the relaxation time. Introducing the rescaled fast time $s = t/\tau$, the system can be rewritten as
\begin{equation}
\begin{cases}
\tau \partial_s T + \nabla \cdot q = 0, \\
\partial_s q = -q - c^2 \nabla T.
\end{cases}
\tag{B1}
\end{equation}

In this formulation, $q$ is identified as a fast variable, while $T$ evolves on a slow time scale. In the limit $\tau \to 0$, the temperature field becomes quasi-static with respect to the fast dynamics, whereas the flux relaxes rapidly toward a constrained manifold. Setting $\tau = 0$ in Eq.~(B1) yields the algebraic constraint
\[
q = -c^2 \nabla T,
\]
which defines a slow invariant manifold in the extended phase space $(T,q)$. Substituting this relation into the conservation law recovers the diffusion equation
\begin{equation}
\partial_t T = \alpha \nabla^2 T, \qquad \alpha = c^2 \tau.
\tag{B2}
\end{equation}

This shows that Fourier’s law does not arise as a regular perturbative correction to hyperbolic transport, but rather as the elimination of fast degrees of freedom governed by the relaxation timescale $\tau$.

The singular nature of this limit is more clearly revealed in Fourier space. For plane-wave modes $e^{ikx + \lambda t}$, the dispersion relation reads
\[
\lambda^2 + \tau^{-1}\lambda + c^2 k^2 = 0,
\]
with eigenvalues
\[
\lambda_{\pm}(k) = -\frac{1}{2\tau} \pm \sqrt{\frac{1}{4\tau^2} - c^2 k^2}.
\]

As $\tau \to 0$, the spectrum separates into two branches:
\[
\lambda_+(k) \to -\alpha k^2, \qquad \lambda_-(k) \to -\tau^{-1}.
\]

The fast mode $\lambda_-$ diverges to $-\infty$, while the slow mode $\lambda_+$ converges to the diffusive eigenvalue. Importantly, this convergence is non-uniform in $k$. For any fixed $\tau > 0$, sufficiently large wavenumbers eventually satisfy $ck \gtrsim (2\tau)^{-1}$, at which point the eigenvalues become complex and wave-like behavior emerges. Consequently, the diffusive dispersion relation is recovered only after the fast branch collapses onto an infinitely damped subspace. From an operator-theoretic perspective, the generator
\[
L(\tau) =
\begin{pmatrix}
0 & \nabla \cdot \\
c^2 \nabla & \tau^{-1} I
\end{pmatrix},
\]
does not converge to the diffusion operator in norm as $\tau \to 0$. Instead, its spectrum undergoes a degenerate collapse in which an entire spectral sector is pushed to infinite decay rate. The limiting parabolic operator therefore fails to approximate the underlying hyperbolic dynamics uniformly in time and wavenumber.

The singular perturbation manifests dynamically through the loss of uniform modal validity. For finite $\tau$, the solution admits a biorthogonal spectral expansion involving both slow and fast modes:
\begin{equation}
\psi(t)
=
a_+(k)\, e^{-\lambda_+(k)t} r_+(k)
+
a_-(k)\, e^{-\lambda_-(k)t} r_-(k).
\tag{B3}
\end{equation}

In the limit $\tau \to 0$, the contribution of the fast mode vanishes only after an initial boundary layer of duration $t \sim \tau$. During this transient regime, the rapid decay of $q$ produces corrections that cannot be captured by a purely diffusive evolution equation.

This fast–slow mismatch implies that diffusion correctly describes the long-time, large-scale behavior but fails to approximate the short-time dynamics, even for arbitrarily small $\tau$. The diffusion limit is therefore singular in time, with distinct asymptotic regimes governed by the same small parameter.

\section{Branch-point topology and half-integer winding of the non-Hermitian spectrum}
As established in Eqs.~(11), (17), and (18), the eigenvalues take the form
\[
\lambda_{\pm}(k) = -\frac{1}{2\tau} \pm \sqrt{\Delta(k)}, \qquad
\Delta(k) = \frac{1}{4\tau^2} - c^2 k^2.
\]

The vanishing of $\Delta(k_c)=0$ defines the exceptional point $k=k_c$, where both eigenvalues and eigenvectors coalesce. Because the eigenvalues are defined on a two-sheeted Riemann surface over complex $k$-space, the EP constitutes a genuine branch point rather than an ordinary degeneracy of a diagonalizable operator.

Locally, writing $k = k_c + z$ with $|z| \ll k_c$, one obtains
\[
\lambda_{\pm}(k)
=
-\frac{1}{2\tau}
\pm \sqrt{-2c^2 k_c z}
+ \mathcal{O}(z^{3/2}),
\]
which explicitly exhibits the square-root singularity underlying the spectral transition.

To characterize this branch-point structure, it is convenient to introduce the spectral gap function
\begin{equation}
g(k) \equiv \lambda_+(k) - \lambda_-(k) = 2\sqrt{\Delta(k)}.
\tag{C1}
\end{equation}

Following standard practice in non-Hermitian spectral analysis, we define a winding number associated with a closed contour $C$ in the complex $k$-plane,
\begin{equation}
W = \frac{1}{2\pi i} \oint_C \partial_k \log g(k)\, dk,
\tag{C2}
\end{equation}
where $C$ encircles the EP once.

Using the local form $g(k) \sim \sqrt{k-k_c}$, one immediately obtains $W = 1/2$. This half-integer value reflects the ramification index of the square-root branch point and encodes the fact that a single encirclement of the EP exchanges the two eigenvalue branches, while two encirclements restore the original sheet.

\end{document}